\documentclass[twocolumn,showpacs,preprintnumbers,amsmath,amssymb]{revtex4}

\usepackage{graphicx}% Include figure files

\begin{document}

\title
{Structure of the Local-field factor of the 2-D electron fluid.
 Possible evidence for correlated scattering of electron pairs.
 }

\author
{
 M.W.C. Dharma-wardana\cite{byline1}}
\author{and Fran\c{c}ois Perrot$^{\ddag}$}
\affiliation{
Institute of Microstructural Sciences,
National Research Council of Canada, Ottawa,Canada. K1A 0R6\\
}
%\date{07-Nov-2002}
\date{2-june-2003}
\begin{abstract}
The static local-field factor (LFF) of the 2-D 
 electron fluid is
calculated {\it nonperturbatively} using a mapping 
to a classical Coulomb fluid 
$\lbrack$Phys. Rev. Lett., {\bf 87}, 206404 (2001)$\rbrack$.
The LFF for the paramagnetic fluid {\it differs markedly}
from 
  perturbation theory where a peak near 2$k_F$ is expected.
 Our
 LFF has a quasi-linear small-k region
leading  to a peak close to 3$k_F$, in agreement with
available quantum Monte Carlo data. The structure in the LFF
and its dependence on the density and temperature
 are interpretted as a signature
of correlated scattering of electron pairs of opposite spin.
The lack of structure at $2k_F$ implies weakened Friedel
oscillations, Kohn anomalies etc.

\end{abstract}
%\vspace{0.5in}
%\hspace{0.5in} see file /usr/people/chandre/tekstuff/xc2d/ms1.tex
%
\maketitle
%
%\section{Introduction.}
{\it Introduction.}---
The physics of the uniform two-dimensional electron
fluid (2DEF) 
 depends crucially on the  ``coupling parameter''
$\Gamma$ = (potential energy)/(kinetic energy).
The $\Gamma$ for the 2DEF at $T=0$ and mean density $n$ is
equal to the mean-disk radius $r_s=(\pi n)^{-1/2}$ per electron.
The parameter $r_s$, the
spin polarization $\zeta$ and the temperature
$T$ are the only variables in this problem.

The response function $\chi(k,\omega)$
is a property of the 2DEF sensitive to exchange-correlation
effects.
It is  expressed
in terms of a reference ''zeroth-order''
 $\chi^0_R(k,\omega)$ and a local-field factor (LFF), 
denoted by $G(k,\omega)$.
\begin{equation}
\label{lffdef}
\chi(k,\omega)=\chi^0_R(k,\omega)/[1-V_k\{1-G(k,\omega)\}\chi^0_R(k,\omega)]
\end{equation}
The LFF is closely related to the vertex function $\Lambda(k,\omega)$.
The static form, $G(k)$, is identical with $G(k,0)$.
As such, considerable effort has been devoted to determining
 $G(k)$, using perturbation theory,  kinetic-equation methods
\cite{rk,stls}, etc.
  A partially analytic, semi-empirical approach
invokes parametrized models constrained
to satisfy sum rules~\cite{iwamoto} which
fit (\cite{tosi,marinescu,teter}) to limited results
obtained from quantum Monte-Carlo (QMC) 
simulations~\cite{moroni}.
While these efforts have considerably extended the available data, it
is still restricted to the fitted $r_s$ regime.  Also, these
fits usually do not invoke any physical model.
 This truly  emphasizes
 the 
 difficulty and delicateness involved in the determination of the LFF.

In this paper we show that the {\it classical}
LFFs, evaluated for a classical Coulomb fluid which is
an approximate mapping of the quantum fluid, agree remarkably
well with the available {\it quantum} data, reproducing the quasi-linear
behaviour in the $k$ region up to about 2 $k_F$, which gets further
extended into a peak structure near 3$k_F$.

	The classical mapping
was discussed in a number of papers~\cite{prl1,prb00,prl2,pd2d} where we
showed that the static properties of the 2D and 3D electron systems,
(or even electron-proton systems~\cite{hyd}), can be calculated
{\it via} an equivalent {\it classical} Coulomb fluid having a
temperature $T_q$ such that it has the same correlation energy
$\epsilon_c$ as the 
quantum system at the physical temperature $T=0$.
The mapping is based on an extension of the classical
Kohn-Sham equation  at $T_q$ so as to mimic the quantum system.
The ``quantum temperature'' $T_q$  
in 2-D was found to be~\cite{prl2},
\begin{equation}
\label{2dmap}
t=T_q/E_F=2/[1+0.86413(r_s^{1/6}-1)^2]
\end{equation}
where $E_F=1/r_s^2$ is the Fermi energy in Hartrees.
At finite $T$, the classical-fluid temperature
$T_{cf}$ is taken to be $(T_q^2+T^2)^{1/2}$, as
discussed in Ref.~\cite{prb00}. The pair-distribution functions (PDFs)
are given by the 
hyper-netted-chain (HNC) equation \cite{hncref}
inclusive of bridge terms.
The HNC equations and the Ornstein-Zernike(OZ) relations are~\cite{hncref}:
\begin{eqnarray}
\label{hnc1}
g_{ij}(r)&=&\exp[-\beta_{cf} \phi_{ij}(r)
+h_{ij}(r)-c_{ij}(r) + B_{ij}(r)]\nonumber\\
 h_{ij}(r) &=& c_{ij}(r)+
\Sigma_s n_s\int d{\bf r}'h_{i,s}
(|{\bf r}-{\bf r}'|)c_{s,j}({\bf r}')
\label{hnc2}
\end{eqnarray}
These involve: (i) the
pair-potential $\phi_{ij}(r)$, (ii) the bridge function
$ B_{ij}(r)$\cite{rosen,yr2d}. The other term, e.g, $c(r)$, is the
 ``direct correlation function''.  These are discussed in
 ref.\cite{prl2,pd2d}, and briefly below.
This  {\it classical} mapping of
 quantum fluids within the HNC
was named  the CHNC.

In effect, although $S(k)$ and related properties
(e.g, $g(r)$, LFF) of a quantum fluid
have to be determined (traditionally) by first evaluating $S(k\,\omega)$,
and then integrating over $\omega$ to obtain $S(k)$, the CHNC
mapping leads directly to good estimates of
$S(k)$, $g(r)$ etc.
In this paper we show {\it numerically} that the classical LFFs
  obtained from CHNC are in remarkable
agreement with QMC data for the available $r_s$ values. This confirms our
basic premise that the static properties of the classical fluid
provide a good approximation to the corresponding
properties of the quantum fluid.

 The 2-D LFFs do {\it not} have the form
indicated by standard perturbation theory~\cite{teter}.
Such  calculations give
LFFs with a ``hump'' at 2$k_F$. 
Here we find that the interactions have extended the quasi-linear
 region and pushed the usual  2$k_F$
 hump towards  $\sim 3k_F$. The limited set of QMC data for
  the 2-D LFF~\cite{moroni} agree with this.
We examine the behaviour of the hump in the LFF, with and with
out the clustering term (bridge term), and  as a function of $r_s$ and $T$,
and find that the
hump in the classical LFF results
from {\i up-down} electron correlations. This suggests
  that correlated scattering
of singlet {\it pairs } may play a role in the
quantum fluid as well, 
since the usual 2$k_F$ anomaly arises from 
scattering of uncorrelated electrons
across the Fermi-disk.

{\it The local-field factor.}---
We consider the static form $G(k)$, defined
 with respect to
a reference ``zeroth-order'' response function. 
The Lindhard function $\chi^0_L(k)$ s often used for this purpose.
However,
%\bibitem{nicklasson}
%Nick, Holos, Vignale
  another natural choice~\cite{nicklasson,teter} is to use
 the ``density-functional''
non-interacting form $\chi(k)^0$ containing the occupation numbers
 corresponding to the
{\it interacting} density. The two choices mainly
affect the large-$k$ behaviour of the LFF \cite{tosi}. 
Thus, for the Lindhard reference used in QMC,
\begin{eqnarray}
\label{asymk}
\lim_{k\to\infty} G(k)&=& C_\infty k + 1-g(0) \\
 C_\infty&=&-(r_s\alpha/2) d[r_s \epsilon_c(r_s)]/dr_s.
\end{eqnarray}
We showed in
Ref.~\cite{prb00} that the LFFs from CHNC have the $1-g(0)$ limit.
The CHNC data and the QMC data can be compared on the same 
footing by removing the asymptotic
linear-$k$ dependence.
 Here $\epsilon_c(r_s)$ is
 in Hartrees/electron,  $k$ is in
units of $k_F$, and $\alpha=1/\surd(2)$.
The remaining term, $1-g(0)$ depends on the estimated ``on-top'' value
of the PDF, a subject discussed by Bulutay et al.~\cite{balutay}
in the context of the CHNC and other methods. 

The CHNC provides a {\it very simple} formula for the LFF, via the
classical-fluid. Unlike in the quantum case,
for a classical fluid, $\chi(k)$ is
directly related to the structure factor.
\begin{equation}
S_{ij}(k)=-(1/\beta)\chi_{ij}(k)/(n_in_j)^{1/2} .
\label{StoChi}
\end{equation}
Hence, taking the paramagnetic case for simplicity, 
\begin{equation}
\label{lfcCS}
V_{c}(k)G(k)=V_{c}(k) -\frac{T_{cf}}{n}
\Bigl\lbrack\frac{1}{S(k)}-\frac{1}{S^0(k)}\Bigr\rbrack .
\end{equation}
Here $T_{cf}$ equals $T_q$
if the physical temperature $T$ = 0.
In CHNC
 the $\chi^0(k)$  and $S^0(k)$ are based on a
Slater determinant and not on the
 Lindhard function.
QMC results use a reference $\chi_L^0$ such that
the LFF contain a kinetic-energy tail, as discussed in 
Eq.~\ref{asymk}

 The $S(k)$ needed in Eq.~\ref{lfcCS}  is
explicitly known from the CHNC calculation.
Alternatively, any other
source of $S(k)$, e.g., QMC, may be used, while $S^0(k)$ for the
2DEG is analytically known. 
The Coulomb potential $V_c(r)$ occurring in Eq.~\ref{lfcCS}
needs explanation.
The Coulomb operator for point-charge electrons is  $1/r$.
However, the classical electron at
the temperature $T_{cf}$ is
localized to within a thermal de Broglie wavelength. Hence, the
effective classical interaction in CHNC
is the ``diffraction
corrected'' form~\cite{prl2}
\begin{eqnarray}
\label{potd}
V_{c}(r)&=&(1/r)[1-e^{-rk_{th}}]\\
V_{c}(k)&=&2\pi[k^{-1}-(k_{th}^2+k^2)^{-1/2}]
\end{eqnarray}
By numerically
 solving the Schrodinger equation
 for a pair
of 2-D electrons in the potential $1/r$ and calculating the electron density
in each normalized state~\cite{pd2d}, we get 
 $$k_{th}/k^0_{th}=1.158T_{cf}^{0.103}$$
 where $T_{cf}$ is in au. Here $k^0_{th}$ is the familiar
3-D form of the de Broglie wavevector, $(2\pi m^* T_q)^{1/2}$,
 where $m^*$ is the effective mass of the
electron pair. We {\it emphasize} that the $G(k)$ can be calculated
only if this modified potential $V_c(k)$ were used. The large-$k$ sum rule
for $G(k)$ is recovered with this $V_c(k)$ which also contains the 
$T_{cf}$ entering into Eq.~\ref{lfcCS}. In effect, the classical equation
 successfully  satisfies the small and large-$k$ sum rules
satisfied by the quantum $G(k)$ itself.

The Coulomb potential $V_c$ becomes large for small $k$, and
 explicit cancellation of $V_{c}$ by the terms in the $1/S-1/S_0$
is desirable in numerical work.
 This can be done by rewriting the structure factors
in terms of the direct correlation functions using the OZ
relations \cite{condmat}.

The bridge term models short-ranged correlations
($k > 2k_F$) in the 2-D LFF. Hard-sphere (\cite{rosen}) or hard-disk
models can be used to obtain an
explicit
form for the bridge function~\cite{prl2,yr2d}.
This gives satisfactory results, esp. for strong coupling
where it is most needed. However, unlike in 3-D classical
fluid-studies (\cite{rosen})
we have not used the bridge function to fit the compressibility sum rule.
The latter is approximately satisfied even without the bridge term.
It was used as a short ranged interaction 
 to
 mimic the back-flow effects~\cite{kwon} of QMC, and comes
into play mainly for $k> 2k_F$.
An improved bridge function would be based directly on the $V_c(r)$
rather than the hard-disk model used here.
The required hard disc diameter $\sigma=2r_s\surd\eta$, where 
$\eta$ is the packing fraction. It is given by
\begin{equation}
\label{eta}
\eta=0.235r_s^{1/3}/[1+0.86413(r_s^{1/6}-1)^2]
\end{equation}
and is based on the dependence required by the Gibbs-Bogoliubov form
of the free energy with respect to a reference fluid.
The bridge term becomes zero when $\eta$ is set to zero, and hence
we can study the LFF with and without cluster terms.
Also, since the $V_{c}(r)$ becomes small for $r\to 0$, the large-$k$
behaviour of the LFF based on the $\eta=0$ bridge function would
 be of interest. Similarly, for small-$k$ (i.e, large-$r$), the
short-ranged correlations are irrelevant and the $\eta=0$ behaviour
is retained.

{\it  Comparison of the CHNC results with QMC data.}---
In Fig.~1 we compare the CHNC results for $r_s=2$, 5, and 10
(with and without
the bridge term), with the QMC results.
The CHNC results for the LFF have {\it not been fitted} to any outside data.
In comparing with QMC, it is necessary to remove the large-$k$
dependence arising from the Lindhard reference (see Eq.~\ref{asymk}).
 The subtraction
of the $C_{\infty} k$ term must be applied asymptotically and this leads to
some arbitrariness in deciding on the ``asymptotic'' regime.
Atwal et al.~\cite{teter} use what they call ''an admittedly {\it ad hoc}''
 scheme in their
Eq. (22).  In Fig.~1 we give the  QMC data (labeled 'qmc-a')
 where the asymptotic term of Davoudi et al.~\cite{tosi}
has been used as a means of removing the $k$-asymptote.
It modifies
even the data points smaller than 2$k_F$. 
After subtraction, the maximum-amplitude LFF occurs for $r_s=10$,
rather than at $r_s=5.41$, as was the case prior to subtraction.

 The LFF from CHNC without the bridge term ($\eta=0$)
shows a large quasi-linear behaviour of the LFF for 
$k$ up to and beyond $2k_F$.  The bridge term 
extends the quasi-linear region and introduces short-range effects
(i.e, for larger-$k$) and producing a
hump, agreeing with QMC data, even though the
QMC $k$-range is limited.
 The CHNC data with and without bridge terms go to the
 CHNC $1-g(0)$ limit for large $k$. The CHNC-bridge LFF has oscillations
in the pre-asymptotic beyond-the-peak region. This is
not seen
in the QMC points. QMC seems to follow the $\eta=0$ curve for large-$k$.
This is probably realistic since $V_{c}(r)$ near $r\to 0$ is
much softer than the hard-disk potential
used by us for the bridge term.

	Thus we see that the classically calculated LFF of the CHNC 
Coulomb fluid provides a good approximation to the QMC generated LFF in
all the available cases. Hence we can use the CHNC to calculate
classical LFFs for other $r_s$ values where QMC data are {\it not} available.
	
%
%\begin{turnpage}
\begin{figure}
\includegraphics*[width=9.0cm, height=10.0cm]{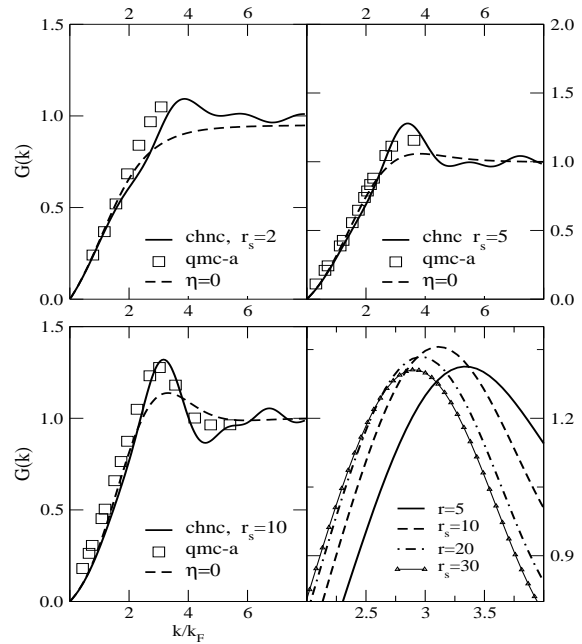}
\caption
{The local-field factors  calculated  from CHNC, with and without ($\eta=0$)
the bridge term.  The QMC data, squares, ref~\cite{moroni} were
extracted from Davoudi et al.\cite{tosi},
 with the asymptotic-$k$
behaviour subtracted out (see text), and labeled ''qmc-a''.
   The lower right panel shows the $\sim 3k_F$ hump structure of
  the LFF  for $r_s$=5,10,20 and 30 obtained from CHNC.
}

\label{lfcfig1}
%fig 1
\end{figure}
%\end{turnpage}

%
The featureless  $2k_F$ region and the appearance of a broad
hump near $3k_F$ are shown for  $r_s$=5,10,20, and 30 in
the lower right panel.  The increased coupling (larger $r_s$) moves the
peak to {\it shorter} wavevectors. The absence of structure near $2k_F$
implies a weakening of Friedel oscillations and Kohn anomalies in 2D.
 
\begin{figure}
\includegraphics*[width=4.0cm, height=4.0cm]{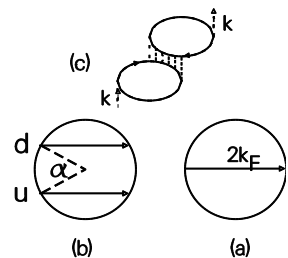}
\caption
{Electron scattering across the Fermi disk. (a) Single-electron
 scattering. (b) Scattering of a correlated u- and d-spin pair.
(c)A typical Feynman graph describing these correlations.
}
\label{fermifig2}
%fig.2
\end{figure}

{\it Discussion}---
	The strong coupling effects in the 2DEG were modeled in CHNC
 via a bridge term limited to short-range interactions ( $k > 2k_F$).
 This term plays
 a role only for anti-parallel spins, i.e., in $g_{12}(r)$. Also,
 available results (not discussed here)
  show that the hump near
 $3k_F$ does not appear for the parallel spin case (the
 antisymmetric LFF), where the $2k_F$ behaviour is similar to that expected
 from low-order perturbation theory. These considerations suggest that the
 structure near $3k_F$ my be a result of {\it correlated} pair processes.
The broadness of the peak suggests that this is not a sharp process.
 The lack of  Pauli exclusion between two opposite-spin electrons and the
 strong coupling would lead to correlated pairs.
Our results show their importance in the classical
fluid which is the CHNC map of the quantum fluid. Hence we make
the hypothesis that such pairs play a role in the quantum liquid as well.
 In Fig.~\ref{fermifig2}(a) an uncorrelated electron
 scatters with an electron
  across the Fermi disk, 
  with a change of momentum 
  $\Delta k$ = $2k_F$. The structure seen in the
  LFFs near
  $2k_F$ in weak coupling arises from such uncorrelated scattering across
  the Fermi disk. Although there is another electron of opposite spin 
  in the same $k_F$-state, it is uncorrelated with
   the scattering electron and takes
  no part in the transition.

   Consider the correlated case, 
   Fig.~\ref{fermifig2} (b), where the up-spin $u$, and down-spin $d$ electron
   are in two states at $k_F$, making an angle $\alpha$ in the Fermi disk.
   If the Coulomb repulsion were absent, the $u$, $d$ pair would 
   occupy the same state with $\alpha=0$. Unlike in the uncorrelated case
   (a), scattering of the correlated $u,d$ pair can occur in a concerted
   manner and would lead to a  $\Delta k$ = $4k_F cos(\alpha/2)$.
	Depending on the correlations existing in the fluid, some optimal
    value of $\alpha$ would be most probable. The maxima in the LFF for
    $r_s$=5, and 20 are
  at $\Delta k/k_F$= 3.34 and $\sim 3$. This mimics the shift in
  In the peak in the $S(k)$ from 2.98 to 2.68 for $r_s$= 5 and 20.
     If these processes are
  to be treated in a diagrammatic theory for the polarization operator, then
  diagrams like Fig.~\ref{fermifig2} (c) are needed.
The peak structure is strongest at the density $r_s=5.41$,
 where correlated singlet-pair effects are most strong. After that
exchange effects begin to counter correlation effects.

\begin{figure}
\includegraphics*[width=8.0cm, height=6.0cm]{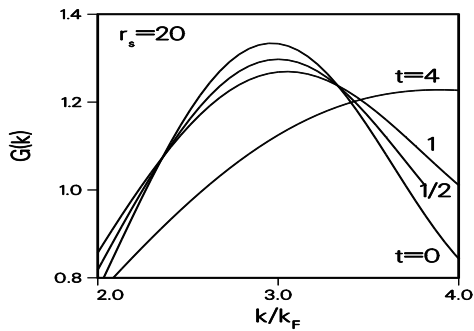}
\caption
{ Temperature ($t=T/T_F$) dependence of the LFF peak at $r_s$=20.
}
\label{figT}
%fig 3
\end{figure}

  At finite temperature, 
  correlated-pair scattering should become weaker and the
 peak position should  shift towards larger $k/k_F$, rather than
 towards 2$k_F$. This is confirmed
 in the finite-$T$ data for $r_s$=20  shown in Fig.~\ref{figT}.

	Finally we remark that the bridge term ( similarly, the
back-flow term in QMC) provides extra pair-interactions which make the
paramagnetic fluid energetically less favourable than the ferromagnetic
phase. Thus the recently proposed {\it para}$\to$ {\it ferro} transition
\cite{atta,pd2d} occurs only if bridge contributions (or, in the QMC
case, back-flow terms) are included in the analysis. That is, there are
no {\it para}$\to$ {\it ferro} transitions in the $\eta=0$ CHNC
 calculation~\cite{balutay}.

%\end{turnpage}
 
 {\it Conclusion}---
 We have shown that the CHNC derived LFF provides a remarkably good
 representation of the quantum simulation results so far available.
 Unusual features of the 2D-LFF not found in the 3D-case, and unexpected from
 perturbation theory, were examined via the CHNC method. 
 The lack of structure near 2$k_F$ and the presence of
 unexpected structure near 3$k_F$ which arises only when cluster-terms
are included in the classical map suggest them to be signatures of
 correlated singlet-pair scattering in the 2-D electron fluid.
 The possibility of such scattering would be very relevant to theories
 of superconductivity in 2-D systems, spintronics and  related 
 topics. The CHNC method thus provides a useful exploratory tool
 for  strongly
 correlated regimes
 inaccessible by standard analytical methods.
  On-line access to our CHNC codes  and more details may
 be obtained at our website\cite{web}.


\begin{references}
\bibitem{condmat}
M. W. C. Dharma-wardana and F. Perrot, condmat/0211127
This paper contains preliminary results from somewhat less accurate
numerical procedures, partly associated with ensuring that the
bridge function was truly short ranged. We find that the best way to ensure
a short-ranged bridge function is to use no bridge function
($\eta=0$) up to $k\sim 2k_F$, and then include a finite $\eta$
for $k > 2k_F$. The cross over is quite smooth, esp. at large coupling.
\end{references}
\end{document}